\documentclass[11pt,twoside]{article}
\usepackage{asp2004}
\usepackage{psfig}
\usepackage{epsf}
\usepackage{graphics}
\usepackage{graphicx}
\usepackage{lscape}
\def\edcomment#1{\iffalse\marginpar{\raggedright\sl#1\/}\else\relax\fi}
\reversemarginpar

\begin{document}
\title{Signatures of Explosion Models for SN ~Ia \& Cosmology}
\author{P. H\"oflich}
\affil{Dept. of Astronomy, University of Texas, Austin, USA}

\begin{abstract}
 Based on detailed models for the progenitors, explosions, light curves (LCs) and spectra,
we discuss signatures of thermonuclear explosions, and the implications for cosmology.
 Consistency is needed to link observables and explosion physics.
 SNe~Ia most probably  result from the explosion of a degenerate CO-White Dwarf (WD) close to the
Chandrasekhar mass. There is strong evidence that most of the WD is burned with an extended
outer layer of  explosive C-burning products (O,Ne,Mg) and very
little C remaining. Overall, the chemical structure is radially stratified.
 This leads to the currently favored delayed detonation model in which a phase
of slow nuclear burning as a deflagration front  is followed by a detonation phase. 
 The importance of pre-conditioning
became obvious.  Within a unified scenario, spherical models allow to understand both
the homogeneity and basic properties of LCs and spectra, and they allow to probe for their diversity
which is a key for high precision
cosmology by SNe~Ia. We emphasize the relation between LC properties and spectra because,
for local SNe~Ia, the diversity becomes apparent the combination of  spectra and LCs whereas, by enlarge, we have to
for high-z objects. We show how we can actually probe the properties of the progenitor,
its environment, and details of the explosion physics. 
  We demonstrate the influence of the metallicity Z on the progenitors, explosion physics and the combined effect
on light curves. By enlarge, a change of  Z causes a shift of along the brightness decline relation because Z
shifts the balance between ${56}Ni$ and non-radioactive isotopes but hardly changes the energetics or the $^{56}Ni$ distribution.
However, the diversity of the progenitors produces  an intrinsic dispersion in B-V which may pose a problem for reddening corrections.
 We discuss the nature of subluminous SN1999by, and how it can be understood 
 in the same framework as 'normal-bright'
SNe~Ia. At the example of SN200du, we show the influence of the progenitor system
and  distribution of isotopes  on light curves. In both objects, we have seen
 clear evidence for some departure
from sphericity probably  due to circumstellar interaction and stellar rotation but the 3D signatures of deflagration
fronts remain an  elusive feature.
\end{abstract}
\thispagestyle{plain}
\section{Explosions, Light Curves and Spectra}
\begin{figure}[h]
\vskip -0.3cm
\includegraphics[width=3.1cm,angle=270]{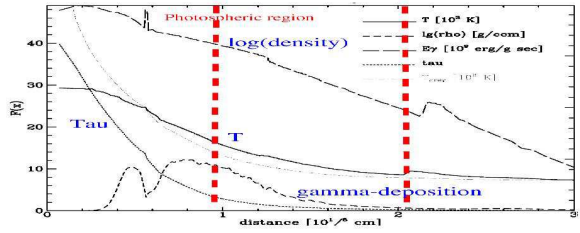}
\vskip -0.2cm
\caption{
Envelopes require radiation hydrodynamical calculations well beyond classical stellar
atmospheres or the commonly used 'light bulb' approach which  assume well defined photospheres with
a black body as inner boundary. The
temperature {\sl T}, energy deposition due to radioactive
decay $E_{\gamma} $, Rosseland optical depth $Tau $(left scale) and density  (right scale)
are given as a function of distance  (in $10^{15}cm$) for a typical SNe~Ia at
15 days after the explosion. For comparison, we give the temperature
 $T_{grey}$ for the grey extended atmosphere.
 The
two dotted, vertical lines indicate the region of spectra formation.
}
\vskip -0.5cm
\label{radhyd}
\end{figure}
   The last decade has witnessed an explosive growth of high-quality
   data for supernovae. Advances in computational methods provided
   new insights into the physics of the objects, and advances in cosmology. 
   Both trends combined provided spectacular results not only for astronomy 
 and the origin  of elements but also for nuclear, high energy and particle physics, and cosmology.
 Further improvements and the quest for the nature of the dark energy requires an increased accuracy
for distance determinations from 10 \% to about 2 to 3 \%
(Weller \& Albrecht 2001) making evolutionary effects with redshift
 a main concern, and a a better understanding of the physics of SNe~Ia a requirement.
There is general agreement that Type Ia Supernovae (SNe~Ia) are the result of a thermonuclear explosion
of a degenerate C/O white dwarf (WD) with a mass close to the Chandrasekhar limit. These scenarios
 allow to reproduce  optical/infrared light curves (LC) and  spectra of
 SNe~Ia reasonably well.
Nowadays, we understand the basic, observational  features.
 SNe~Ia appear rather homogeneous because
nuclear physics determines the structure of the WD,  the explosion, light curves and spectra:
 (1) the WD  is supported by degenerate electron pressure, (2) the
 total energy production during the explosion is given by the
release of thermonuclear energy,  and (3) the light curves are powered by the radioactive
decay of $^{56}Ni$ produced during the explosion
To first order,
the outcome hardly depends on details of the physics, the scenario, or the progenitor
("stellar amnesia"). Homogeneity of SNe~Ia does not (!) imply a unique 
scenario, and it took the revolution in observational methods with respect to time and wavelength 
coverage to reveal differences and expose the diversity of within SNe~Ia.
 For recent reviews see Branch (1999) and H\"oflich et al. 2003).
\begin{figure}[h]
\includegraphics[width=4.3cm,clip=,angle=270]{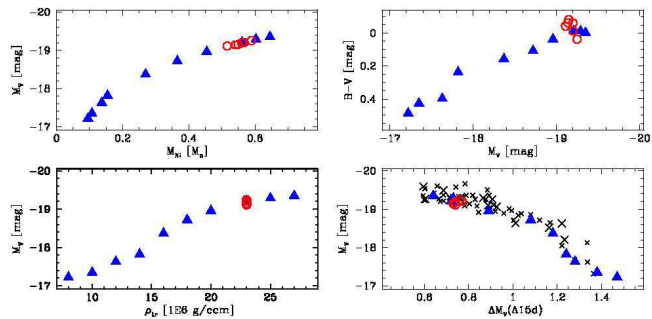}
\vskip -0.30cm
\caption{
  Light curve properties of DD models with various transition densities (triangles,
H\"oflich et al. 2002), and with a WD originating from progenitors with main sequence masses $M_{MS}$ between  1.5 to 7 $M_\odot$ and 0
to solar metallicity (open circles, Dominguez et al. 2002).
 We give the maximum brightness $M_V$ as a function of ${56}Ni$ production (upper left) and the transition density $\rho_{tr}$(lower left),
the dependence  of the intrinsic color B-V($M_V$) on $M_V$  (upper right), and $M_V$  as a function of the post-maximum decline $\Delta M_V (15d)$(V)
in $V$ over 15 days (lower left). In addition, we plot observed $\Delta M_V(15d)$ of 66 well-observed SNe~Ia (Phillips, 2003). For the transformation between $\Delta M_V (15d)$(B) to
$\Delta M_V (15d)$(V), see Garnavich et al. (2004).
}
\vskip -0.2cm
\label{chem}
\end{figure}
\begin{figure}[h]
\vskip -0.2cm
\includegraphics[width=2.7cm,angle=270]{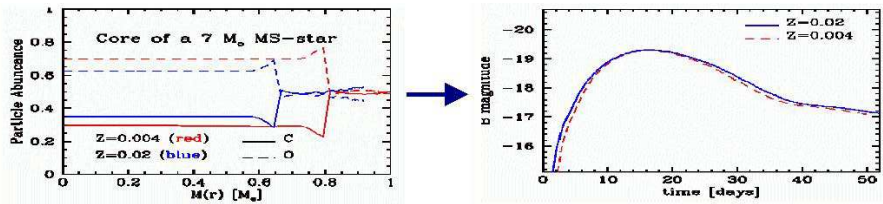}
\caption{
 Influence of the metallicity Z
  on the B and V light curves for a progenitor star of 7 $M_\odot$ on the
main sequence based on a delayed detonation model.
 For the progenitor, Z[O/Fe] is  taken according to  Argast et al. 2000.
}
\vskip -0.3cm
\label{prog}
\end{figure}
\begin{figure}[h]
\vskip -0.4cm
\includegraphics[width=5.4cm,angle=90]{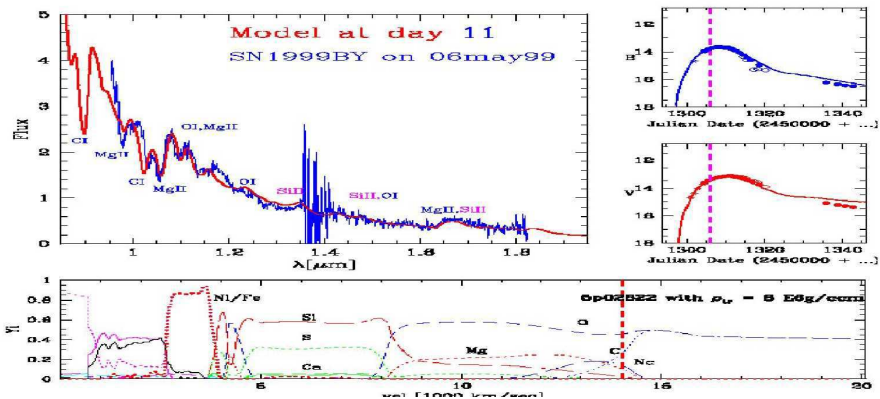}
\caption{
 Analysis of the subluminous SN1999by.
 We show IR-spectra at day 11 (upper left),
the B and V light curves  (right plots), and the chemical structure
 (lower panel).
The explosion  and  evolution of the spectra are
calculated self-consistently with the only free parameters being the initial
structure of the exploding White Dwarf, and a parameterized description
of the nuclear burning front.
Without further tuning, the spectra and
their evolution with time can be reproduced.
 Up to maximum light,the spectra are formed in layers of explosive C-burning, followed by
the layers of incomplete Si burning. The long duration of these phases
provides a lower limit for the initial WD mass which is close to the Chandrasekhar mass.
 Unburned material is restricted to the high velocity regime, i.e. the very outer layers.
This strongly supports DD models for both the normal bright and subluminous SNe~Ia.
}
\vskip -0.2cm
\label{sn99by}
\end{figure}
\begin{figure}[h]
\vskip -0.0cm
\includegraphics[width=5.0cm,angle=270]{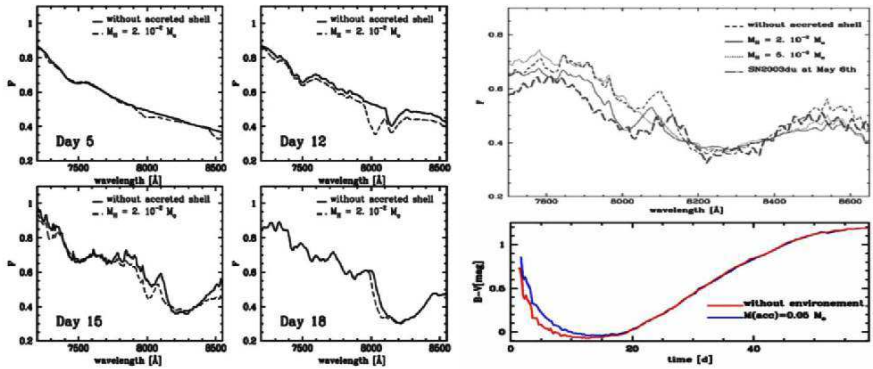}
\caption{
 High velocity CaII feature as tell-tail for interaction
within the progenitor system.
We show the CaII IR feature observed in SN~2003du on May 6rd in comparison with theoretical models
at about 15 days after the explosion (upper right), and its evolution with time (left).
 The calculations are based on a delayed detonation model which interacted with a nearby,  shells
 of 0.02  $M_\odot$  and solar composition during the early phase of the explosion.
 The dominant signature of this interaction is the appearance of a persistent, secondary, high velocity Ca II feature.
Without ongoing interaction, no H or He lines are detectable.
 Note that, even without a shell, a secondary Ca II feature can be seen for a period of 2 to 3 days
during the phase when Ca III recombines to Ca II emphasizing the importance of a good time coverage
for the observations. Nearby shells mainly change early time LCs (lower right) due to blocking by Thomson optical depth
in the shell. In contrast,
ongoing interaction will change the late time luminosities (from Gerardy et al. 2004).
}
\vskip -0.0cm
\label{ca}
\end{figure}
\begin{figure}[h]
\vskip -0.0cm
\includegraphics[width=4.8cm,angle=270]{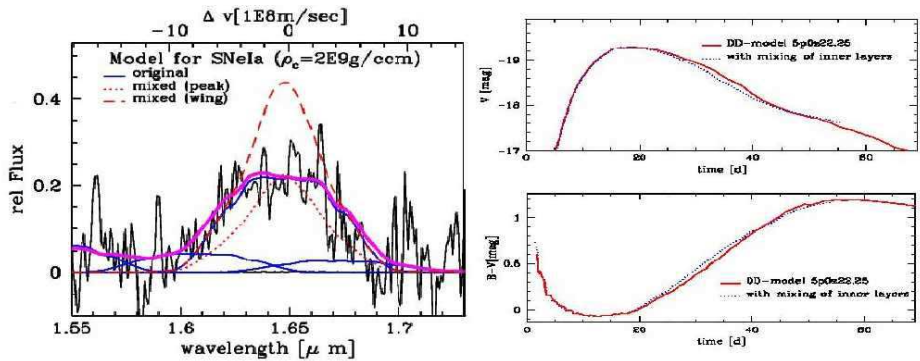}
\caption{ Late time IR-spectra a probe of
the isotopic structure and its influence
on the optical light curves. On the left, we give  the
NIR spectrum of the "Branch-normal"  SN 2003du on Febr. 27, 2004 with Subaru, in comparison with
the theoretical line profiles (solid thick line) based on a DD model with a central region
consisting of non-radioactive iron group elements. 
 In addition, the individual components of  the forbidden [Fe~II] transition at 1.644 $\mu m$ 
are given for the original delayed detonation model (solid) and mixed chemistry (light) 
normalized to the maximum line flux (dotted) and the wings (dashed), respectively.
 Mixing of the inner iron-rich layers of ${56}Ni$ and stable isotopes (Fig. 4) is to be expected 
 from current 3D models during the deflagration phase which is dominated by RT instabilities, and 
would produce round profiles which seem to be at odds with the observations. Possible explanations may be that
 small-scale, pre-existing velocity fields are important  for the propagation of nuclear flames.
On the right, the visual light curve and $(B-V)$ are given for the same delayed detonation model but with
and without mixing of the inner layers. Differences in V and B-V are $ \approx $0.2$^m$ and $0.05^m$,
 respectively. In effect, mixing redistributes $^{56}Ni$ from the outer to the inner layers which decreases
the photospheric heating at about maximum light but increases the $\gamma$-trapping later on (from H\"oflich et al. 2004).
}
\vskip -0.0cm
\label{isotopes}
\end{figure}
\begin{figure}[h]
\vskip 1.0cm \includegraphics[width=5.5cm,angle=270]{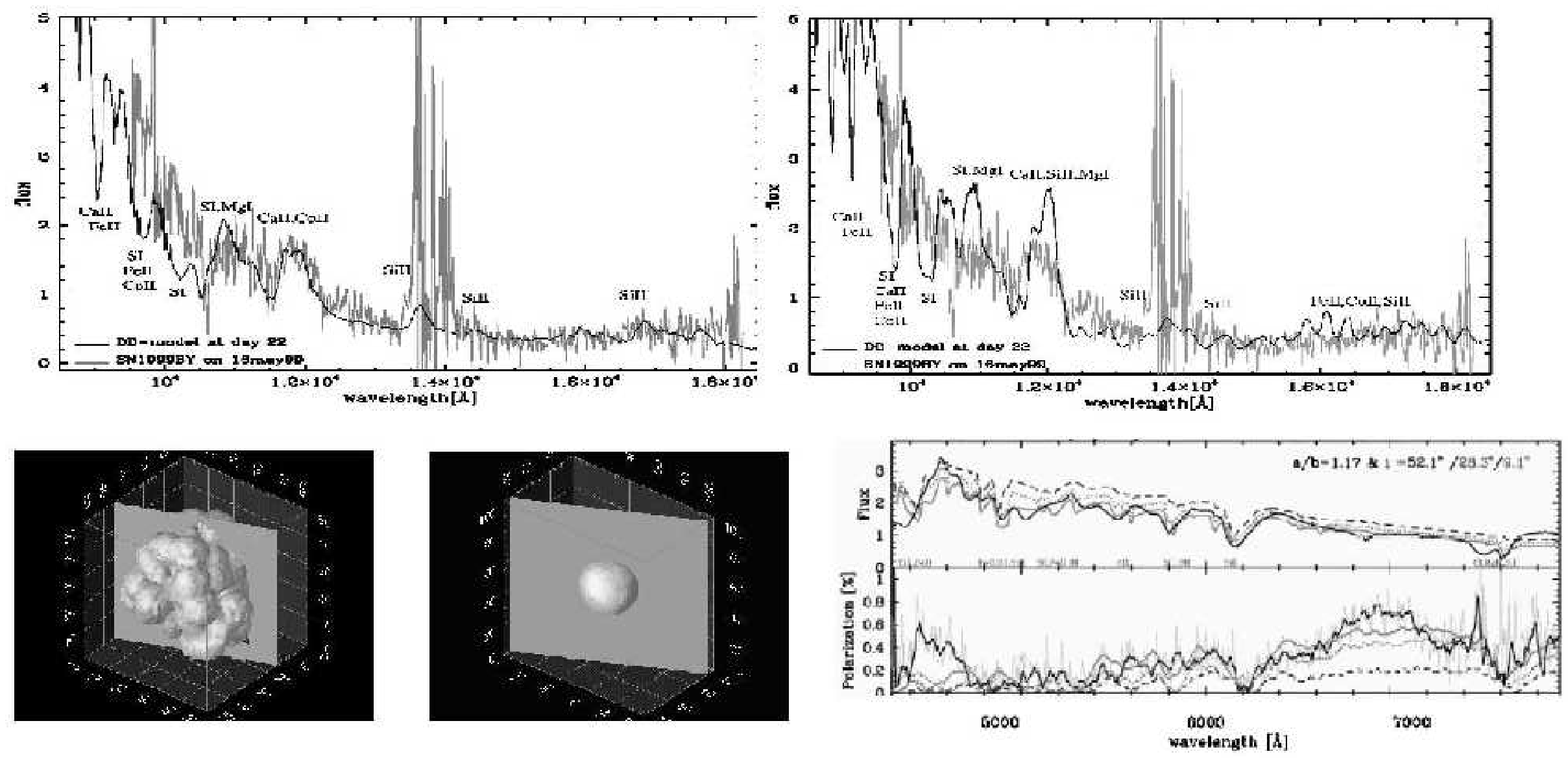}
\caption{
 Analysis of 3D effects in the subluminous SN1999by based on  flux and polarization
data. {\bf Upper panel:}
Comparison of the  NIR spectrum on
May 16 (left) with a spherical, subluminous delayed detonation model without and with mixing
as expected from  detailed 3-D deflagration models (Khokhlov, 2001) which does destroy any fit due to
excitation of intermediate mass elements (S,Si), and the absorption by iron-group elements.
{\bf Lower, left panel:} Energy deposition by $\gamma $-rays at day 1 (left) and 23 (right) for a 3D deflagration
model based on our full 3-D MC gamma ray transport.
At about day 23, the energy deposition
is not confined to the radioactive $^{56}Ni$ ruling out clumpiness as a solution
to the excessive excitation of S and Si lines.
{\bf Lower, right panel:} Optical flux and polarization spectra
 at day 15 after the explosion for the subluminous 3-D delayed-detonation model in comparison with the
SN1999by at about  maximum light.
In the observations, the polarization angle is constant indicating  rotational symmetry
 of the envelope, and
an axis ratio A/B
of 1.17 (from H\"oflich et al. 2002, H\"oflich 2002,  Howell et al. 2001).
}
\vskip -0.0cm
\label{3D}
\end{figure}

 A detailed discussion is well beyond the scope of this contribution. We want to
illustrate some aspects and, for details, refer to the original publications.
 
\noindent
- Consistency is required to link the observable and the progenitor system, progenitor and
explosion physics. By the physical conditions, consistency is also demanded for the
treatment of hydrodynamics, rate equations, and  radiation transport
 (Fig.1).
Density structures require detailed hydrodynamics,
low densities cause strong non-LTE effects throughout the entire envelopes and the radiation field differs from a
black body, chemical profiles are depth
dependent,  energy source and sink terms due to hydrodynamical effects and radioactive
decays  dominate throughout the photon decoupling region, and all physical properties are
time-dependent because the energy diffusion time scales are comparable to the hydrodynamical expansion time
scale (H\"oflich, 1995).
Our approach  significantly reduces the number of free parameters,
namely the initial structure of the progenitor, the accretion rate on the WD,
and the description of the nuclear burning front. The light curves and spectral 
evolution follow directly from the explosion model without any further tuning.

\noindent
- The best current explosion scenario  is  the delayed detonation (DD) model (Khokhlov 1991).
 These models  reproduce reasonable well the color evolution of optical
and IR-LC, and the spectral evolution of both normal bright (H\"oflich 1995) and subluminous SNe
(Fig. 4, H\"oflich et al. 2002) within a unified scenario. By enlarge, DD models produce radially stratified
chemical structures for elements of explosive carbon, oxygen and Si burning in agreement with the
observations from the optical and near IR (e.g. Fig. 4, Barbon et al. 1989, Wheeler et al. 1998, Marion et al. 2003)
 The brightness decline relation (Phillips 1993) can be understood as an opacity effect 
 (Fig. 3 , H\"oflich et al. 1996, Maeda  et al. 2003) including the color evolution in
B-V and the brightness range (Fig. 2 H\"oflich et al. 2002) by varying a single, free parameter for
the amount of burning when the deflagration turns into a detonation which determines the pre-expansion of the WD. In spherical models,
this quantity is often parameterized by a transition density $\rho_{tr}$. Other properties produce a dispersion. E.g.
changes in the progenitor, i.e. in main sequence mass $M_{MS}$ and Z, alter the explosion energy and   $^{56}Ni$ mass, respectively,
  but, by enlarge, along the $M_V(^{56}Ni)$ and $\Delta M_V (15d)$
relations (H\"oflich et al. 1998, 2001).
 However, the diversity in  progenitors cause an intrinsic dispersion in B-V of $0.1^m$ which, in practice, seriously hampers the
corrections by interstellar reddening.

\noindent
- The IR is a primary tool to study intermediate mass elements, and  spectropolarimetry and line profiles allow
to probe for 3-D signatures
(H\"oflich 1991, Wang et al. 1997, Bowers et al. 1997,  Wheeler et al. 1998, Howell et al. 2001, H\"oflich et al.2002).

\noindent
- Our models allow to study particular physics and relations to
the properties of the progenitors in 'isolation' such as signatures of the progenitor system (Fig. 5, Gerardy et al. 2004) and
the distribution of neutron rich isotopes (Fig. 6, H\"oflich et al. 2004). In extensive studies, we investigated the influence
of the central density of the WD, the ignition process, and metallicities (which are expected to change with time)
(H\"oflich et al. 1998, 2001, Dominguez et al. 2002, see also Timmes et al. 2003).
 Pre-conditioning of the WD is a key
to understand the diversity of SNeIa. Metallicity will effect the statistical
sample because it changes the life times and radii of stars and, consequently, the binary system with mass overflow,
and the physics of individual objects.
A change of metallicity or progenitor MS mass effects both the progenitor evolution, WD structure
and the observable LC shapes simultaneously, and the combined effect must be considered when comparing to observations.
 As can be seen from  Fig. 3.
 The change of the initial Fe changes the core He
burning during the stellar evolution which determines the C/O ratio, whereas the $^{22}Ne $ directly effects
the explosive nuclear burning by shifting the Nuclear Statistical Equilibrium slightly away from $^{56}Ni$. 
The former effect alters the  energetics of the explosion (explosion energy) whereas the latter changes the
energy gain by
radioactive decay of $^{56}Ni$ which powers the LCs.
  For low metallicities, the influence
of the energetics dominates the changes in the LC shape. It causes a change $\Delta t_{rise}$
of  the rise-time for a given decline ratio and,
thus, an offset $\Delta M$ in the brightness decline relation by $\Delta M \approx 0.1 \times \Delta (t_{rise})$.

These results do not (!) imply that all processes involved are spherical in nature but they are 
a result of the 'stellar amnesia' mentioned above, i.e. the properties depend mostly on integral 
quantities and basic nuclear physics.
 Despite the successes of spherical delayed detonation models, it becomes increasingly obvious
that multidimensional effects are important towards a better understanding of SNe~Ia.

\noindent
- Asymmetries in a SN~Ia may be present for a variety  of reasons. Within the framework of the explosion
of a massive WD, reasons and all seems to be realized. Reasons include instabilities in the burning
front (Khokhlov 1995, 2001, Reinecke et al. 2002 Gamezo et al. 2003), namely the deflagration phase, off-center detonations or the transitions from deflagration
to detonation (Livne 1999),  rapid rotation of the WD, and interaction with the accretion
disk and the companion star. Whereas we have evidence for the existence of the latter effects,
 direct evidence is still elusive  for the RT instabilities or, more precisely, such evidence has not been seen
although it was expected (Figs. 4 \& 6).

 Both from the remnants and the lack of strong polarization in thermonuclear supernovae, limited scattering in brightness,
we can conclude  that the envelopes have  rather small global deviations from spherical geometry of the order
of 5 to 20 \%.  For cosmology,
deviations from sphericity will introduce a directional dependence of the luminosity of the order of about 4-6 \%.

\end{document}